# Architecture for Cooperative Prefetching in P2P Video-on- Demand System

Ubaid Abbasi and Toufik Ahmed

CNRS LaBRI Lab. – University of Bordeaux, France

351, Cours de la Libération

Talence Cedex, France

{abbasi, tad} @labri.fr

*ABSTRACT*

*Most P2P VoD schemes focused on service architectures and overlays optimization without considering segments rarity and the performance of prefetching strategies. As a result, they cannot better support VCR-oriented service in heterogeneous environment having clients using free VCR controls. Despite the remarkable popularity in VoD systems, there exist no prior work that studies the performance gap between different prefetching strategies. In this paper, we analyze and understand the performance of different prefetching strategies. Our analytical characterization brings us not only a better understanding of several fundamental tradeoffs in prefetching strategies, but also important insights on the design of P2P VoD system. On the basis of this analysis, we finally proposed a cooperative prefetching strategy called "cooching". In this strategy, the requested segments in VCR interactivities are prefetched into session beforehand using the information collected through gossips. We evaluate our strategy through extensive simulations. The results indicate that the proposed strategy outperforms the existing prefetching mechanisms.*

## 1. INTRODUCTION

Over the past few years, there has been considerable research in the use of P2P network for the distribution of both live and stored video. Recently, we have witnessing the emergence of a new class of P2P applications such as P2P audio and video streaming. Video streaming supports a limited number of simultaneous users and consume more network bandwidth as compared to other internet applications. On the other hand, the spectacular development in Peer-to-Peer (P2P) networks presents great scalability and support large number of users worldwide. Moreover, P2P networks are automatic systems with self-sharing capability. Each peer consumes and forwards the content to other at the same time, thereby removing the bottleneck of central servers and reducing the cost [1].

Several P2P streaming systems have been deployed to provide live or on-demand video streaming services over the Internet. In live streaming systems, the source server broadcasts the content and all the clients play the content at the same progress. The data chunks must arrive in sequence, so that they can be played at receiver with minimum delay. On the other hand, VoD is an interactive multimedia service in which users enjoys the video with completely free choices due to the availability of VCR-like controls (i.e., forward, backward, resume). The increased link capacity offered to Internet users, has led in popularity of VoD services like YouTube [2], Daily motion [3] and other over the top (OTT) video services. The provision of VoD service to large population of users requires a significant amount of bandwidth and causes the problem of scalability due to terminal and access network heterogeneities. For instance, the bandwidth provisioning costs of YouTube servers are estimated at $1M per day [2]. Therefore Peer-to-Peer (P2P) has emerged as an alternative approach for providing highly scalable streaming services.

Most of the existing work on P2P-based VoD systems has made an implicit assumption that a user who has joined a streaming session will keep on watching till it leaves or fails the session. The careful analysis of real user viewing logs negates the above assumption and reveals the fact that users don't watch the video from beginning to end [4]. In addition, the freedom of content choice leads to more short view sessions. It has been observed that 30% of view sessions are less than 5 minutes. This observation indicates that users join or leave the overlay frequently in P2P VoD. Secondly, random seeks are common and unpredictable in VoD [5]. This results in increase latency and causes excessive stress on streaming server.

Data Prefetching has been proposed as a technique for reducing the access latency. In this technique, peers prefetch and store various portions of the streaming media ahead of their playing position as shown in Fig. 1. In the figure, a small portion of content (urgent downloading target) near to playhead positions are considered more important, and therefore they have been given higher priority. Although, data prefetching requires additional bandwidth and storage but considering the increasing bandwidth on network and storage capability on peers, it actually offers a more desirable tradeoff between quality and cost.

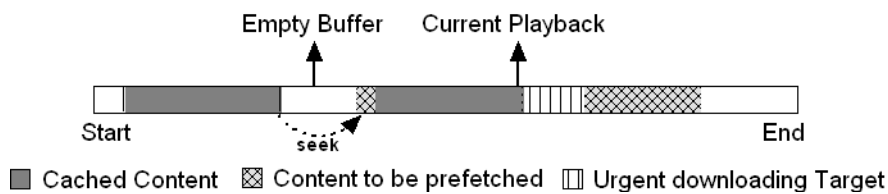

Figure 1: Prefetching, caching and urgent downloading

It is important to differentiate between caching and prefetching. The basic idea behind the caching techniques is that if two clients, request the same video at different times, the server may serve the latter one using the data, which is already cached on behalf of the former one. Thus, the referenced video is read from the source only once, while the system can support in this way several simultaneous displays of the video. In other words, data caching is reactive in nature. On the other hand, data prefetching is a proactive policy which minimizes the seek-to-transfer time ratio. Instead of reading a single data block more are fetched into cache. An effective prefetching strategy must consider the following three factors:

- *Coverage:* The fraction of memory occupied by the prefetched content. The buffer should be distributed efficiently for prefetching video content.
- *Accuracy:* The prefetched content actually used by the peer. The term also referred as *Hit Ratio* and it shows the effectiveness of prefetching strategy.
- *Timeliness:* Data must be available to peers before it is needed but not so early that it is discarded without used. This property is called timeliness.

In this paper, we examine different prefetching strategies and discuss the design tradeoffs involved when implementing these strategies. To the best of our knowledge, this is the first work which provides a performance comparison between different prefetching strategies in P2P VoD systems. On the basis of this analysis, we propose a new prefetching strategy to overcome the discrepancies in existing prefetching strategies. Moreover, our paper focuses on single video approach in which peers only redistributes the video, they are currently playing. The single video approach provides better performances if a large fraction of the requests are for a relatively small number of videos [9].

The rest of the paper is organized in different sections as followed. A brief related work and motivation is presented in section 2. In section 3, we discuss the proposed architecture for cooperative prefetching strategy. Section 4 illustrates the performance evaluation and section 5 presents a brief conclusion while highlighting some of the future perspectives.

## 2. RELATED WORKS

In the past few years, several researches have been proposed for multimedia caching and prefetching. Cheng et al [6] proposed a scheme in which, scheduler adaptively fetches chunks to buffer periodically. The scheduler determines the number of chunks to be fetched after analyzing peer's capacity. Rejaie et al [7] proposed a proxy caching mechanism for layered-encoded streams in internet to maximize the delivered quality of popular streams among interested clients. Although, prefetching of continuous media is discussed but there is no concrete solution for prefetching content in random and unpredictable environment. The optimal off-line prefetching algorithm and a heuristic prefetching algorithm were proposed in [13]. It is shown that performance of layered video can be improved by applying appropriate prefetching policies and prefetching provides higher gain to the layered stream.

### 2.1. Prefetching Techniques for VoD Systems

In this sub-section we analyze the performance of VoD systems without using any prefetching ("no prefetching) strategy. After that we discuss three most common techniques used for data prefetching in VoD systems. These are *random prefetching* [6], popularity *aware prefetching* [5] and *data mining based prefetching* [8].

The peer in a VoD system can prefetches the content in many different ways. We first consider the simplest scheme, which is called no-prefetching [9]. Under this technique, each peer obtained the content at streaming rate and don't prefetch the content for seek operations. The seek operation by a user results in increased latency due to runtime prefetching. Hit ratio is zero in this technique because content are not available in local cache. This technique increases server stress because neighboring peers don't have the desired content and most requests are satisfied by the server. Thus user's playback experience will be degraded seriously.

The random prefetching [6] is used to prefetch the data in local cache before a seek operation is carried out. Rather than waiting for a cache miss to perform a prefetch, random prefetching anticipates such misses and issues a fetch to local cache in advance. Every peers caches a recent few seconds of the data, which is replaced by new one as sliding window proceeds. If a peer wants to prefetch data, it will request the peers having minimum playhead distance with it. The playhead distance is the time difference of current playing position between two peers. Data chunks are prefetched randomly from other peers. The scheduler is responsible for prefetching segments in periodic intervals. If a segment is not available in neighborhood, it can be either requested from server or far neighbors. A prefetched segment, not consumed for a certain period of time will be discarded and a newly fetched segment is placed in buffer. Although, prefetching of data is considered in random prefetching but prefetching of data in unpredictable user behaviors have not been addressed. As a result more useless segments occupy the local cache. Furthermore, the unavailability of useful content increases access latency in random prefetching.

The popularity aware prefetching [5] uses segments access probability for prefetching the content. In this technique logs are maintained by a management server (also called tracker) regarding users access pattern. The statistics gathered on user requests are used to determine the optimal number and placement of replicates for each individual video file. Each peer records its VCR operations information and sends it to tracker or management server after periodic interval. The management server performs the accounting functions on the statistics provided by all peers playing a particular media file. The obtained list of popular content is then distributed among peers. The scheduler of each peer requests for those popular content closest to its current playback position. The popularity aware prefetching techniques improves hit ratio by considering user's access patterns, however large computation are required to be performed by management server for extracting the list of popular content. The periodic exchange of seek operation's information with management server results in additional overhead.

State maintenance and data mining [8] are used as another technique for prefetching more desirable segments of video. Each segment is identified by a unique segment number. Whenever a segment is played, its unique number is saved in a list by each peer. This playback history is exchanged among a set of peers (neighboring peers) which share the closest playhead positions. This playback history provides peers a data set for performing data mining operations. Once a peer received the list of segment being played by neighbors (or peers in same session), data mining is used to find the segments closely related to the current segments. Unlike popularity aware prefetching, each peer performs data mining operations locally instead of central management server. For that purpose, association rule mining is used to find maximum occurring segments with respect to current position.

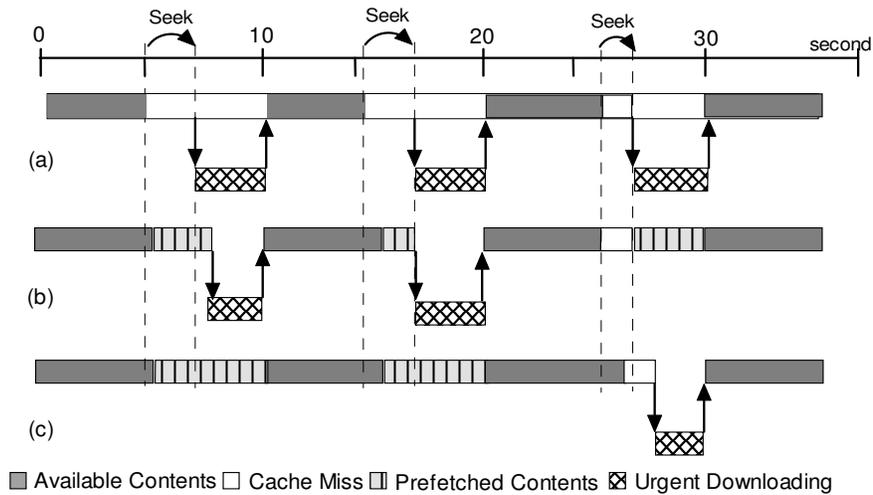

Figure 2: Execution Diagram a) No prefetching, b) Random Prefetching, c) Popularity-aware and data mining based prefetching.

This technique proves greater hit ratio comparatively with no prefetching and random prefetching however, computational cost increases due to data mining operations on obtained list of records. Propagation delay and memory cost for storing list of records is also questionable. Fig. 2 shows the snapshot of buffer window for different prefetching techniques.

## 2.2. Prefetching Techniques Analysis

We defined two types of hit ratios regarding VCR operations. $HR_r$ is the "relative hit ratio", defined as "*the number of prefetching request satisfied locally*". This means that the new segment pointed as a result of seek operation was already available in cache. $HR_g$ is the "global hit ratio", defined as the number of prefetching requests satisfied by requesting segments from neighboring peers. This means that the segment pointed as a result of seek operation was not available, and therefore a prefetch request has been made to obtain the segment from other peers in same session.

Let us suppose, each peer has a memory to store $s$ segments. Note that this is the maximum number of segments that can be prefetched before the occurrence of VCR operations. Let $S$ be the maximum number of segments that can be prefetched by neighboring peers. We will denote $V$ = {set of segments that can be requested using VCR control}. Obviously this set contains either those segments which are not played or those segments which are played but removed from memory later on. We will denote $V_i$ as the set of segment prefetched by peer $i$. let $P_i$ be the probability that the requested segment (as a result of VCR control) exists in $V_i$.

In case of data-mining based prefetching mechanism for peer $i$, we have:

$$HR_r = \frac{P_i \times s}{V_i}, HR_r + HR_g = \frac{P_i \times S}{V_i} \qquad (1)$$

For no prefetching, there is no possibility that the requested segments exists in the set of available prefetched segments. Thus for no prefetching:

$$HR_r = 0, \ HR_r + HR_g = \min imum \qquad (2)$$

Similarly for random prefetching:

$$HR_r = \frac{s}{V}, HR_r + HR_g = \frac{S}{V} \qquad (3)$$

For popularity aware based prefetching we have:

$$HR_r = \frac{P_i \times s}{V}, \ HR_r + HR_g = \frac{P_i \times S}{V} \qquad (4)$$

The above equations show that both data-mining based prefetching and popularity-aware prefetching had better hit ratio comparatively. However popularity-aware prefetching is based on a larger data set, which didn't represent the user's behavior in a particular session. On the other hand data mining based techniques prefetched the popular content consumed by peers having closest playhead distance.

## 3. ARCHITECTURE

We used a tree based overlay structure similar to P2Cast in which peers are organized into different session according to their arrival time. P2Cast peers not only receive the requested stream from parent peer but also contribute to the overall VoD service by forwarding the stream to other peers and caching and serving the initial part of the stream. The detail architecture for cooperative prefetching strategy is explained below.

### 3.1. Data Prefetching Strategy

The drawbacks of different prefetching techniques discussed earlier in section 2 help us to model our problem and propose a better solution. Our aim is to maximize the throughput of the system, by prefetching maximum number of blocks in local cache with minimum playback latency.

### 3.1.1. The Solution: Cooperative Prefetching Technique

We observed a tradeoff between user behavior and overhead in different aforementioned prefetching techniques. Moreover either the management server or each peer has to perform necessary computation which increases computational overhead. To overcome the above mentioned problem, we proposed a new prefetching technique called "*cooperative prefetching technique*". In this technique, each peer maintains the record of playback segments by other peers in the same session. This information is obtained through gossiping. Once the state information are collected from all peers (in same session), each peer creates a table of available segments in that particular session. Fig. 3 shows the state information table received by a particular peer *i*.

| Peer ID | Buffer Map Records |
|---|---|
| J | 1,3,4,5,7,8,9,12 |
| K | 2,3,4,8,9,11,12,13 |
| L | 7,8,9,12,13,14,15,16,17 |
| M | 1,4,5,6,7,13,14,15,20 |
| N | 5,6,8,9,13,14,15,16,17 |
| P | 1,2,3,4,5,6,7,8,11,12 |
| Q | 1,2,4,5,6,7,11,12,14,15 |

State Information of Peer *i*

⇒ | 1 | 2 | 3 | 4 | 5 | 6 | 7 | 8 | 9 | 11 | 12 | 13 | 14 | 15 | 16 | 17 | 20 |

Figure 3: Removing Segment Redundancy

Each peer performs the necessary computation to remove redundancy and creates a list of available non redundant segments in the session. The peer then requests for a segment near to its playhead position, which didn't exist in that session. In case of Fig. 3 missing segments like 10, 18, and 19 would be requested from *shortcut neighbors* (peers in other session) depending on current playhead position. Thus the segment request is broadcasted to other session. As a result, those rare segments are obtained from other session using the *shortcut neighbor list* that didn't exist in the current session. Later on, if a seek operation is carried out and the segment is available in the same session, it will take less time to acquire it from neighbor peers instead of server or far peers. If there is no response from shortcut neighbors, the desired segment is requested from server as a last resort. It is important to note that each peer also prefetch the segments near to its playhead position as an *urgent downloading target*. In our case each peer prefetch the next 20 seconds of video segments as urgent downloading target. Apart from these segments, remaining segments are prefetched using cooperative prefetching strategy. The algorithm for cooperative prefetching is presented in Fig. 4.

When a peer requests for missing segments from other peers (within or outside the session), it can receive multiple response. The selection of appropriate peer for provision of the desired segment is also crucial. In cooperative prefetching each peer performs local scheduling by using the local information available at each node. Our approach is to utilize the previous history of traffic from the peers. Let $W_{ki}$ represents the estimated rate at which a certain peer *k* can deliver to peer *i*. Let $f_{ki}^P$ denotes the total number of segments received by peer *i* from a certain peer *k*. For each request period we use the average number of segments received by node *i* in the last *P* periods.

$$W_{ki} = \frac{\sum_{T=p-P+1}^{p} f_{ki}^T}{P} \qquad (5)$$

The above equation shows that each peer keeps track of the number of segments received by different peers in last *P* periods. By using this information each peer selects an appropriate peer for reception of video segments.

```
    Segment Select ()      //Find the next segment to prefetch
    {
      Find the segments S_i that didn't exist in buffermap
           Return S;      //S is the desired segment;
    }
    Void Prefetch ()       // The function to do prefetching
      {
         While (prefetching set is not empty)
           {
            segment S = select ();
            Broadcast(S); //Broadcast the prefetching of segment
           If (segment S is cached by a peer in same session)

           //situation where same segment is also requested by some other peer
               {
                 Download segment S;
                 Remove the segment S from prefetching list;
               }
           else if (segment S is located on a remote peer P)
               {
                 Connect with the peer P;
                 Download Segment S;
                 Remove segment S from the prefetching set;
               }
           else // when timeout expires
               {
                Send the segment request to server;
                Connect with server;
                Download segment S;
                Remove segment S from the prefetching set;
                }
           }}}
```

Figure 4: Cooperative Prefetching Algorithm

Our strategy selects priority packets near to playhead position as *urgent downloading target*. Furthermore, our strategy focused on segments that are missing in current session. This technique ensures that maximum numbers of segments are brought to the session in order to minimize the delay that occurs during VCR operations. The local scheduling ensures that only a single peer with higher contribution is selected for retrieving missing segments.

The peers in our system execute periodical workflows to leverage the design essentials above. When a peer is streaming video, it maintains a record of available segments. Peers exchange their records through gossips, so that the states of peers are efficiently propagated and updated throughout the networks. Based on the information collected, each peer conducts cooperative prefetching to maximize the resource availability in a session. As a result, the response latency for any VCR control is restricted to a fairly low level.

## 4. PERFORMANCE EVALUATION

This section describes the performance evaluation of cooperative prefetching strategy for different parameters using NS2 simulations [10]. We performed intensive simulations to study the performance of cooperative prefetching strategy against existing prefetching strategies discussed earlier in section 2.

*Network Topology:* We used the BRITE [12] universal topology generator in the top-down hierarchical mode to map the physical network. The network topology consists of autonomous system (AS) and fixed number of routers. All AS are assumed to be in the Transit-Stub manner. Peers are attached to the routers randomly. The delay on the access links is randomly selected between 5 to 10ms. In order to support the streaming service, the peers should be able to download with rate at least the rate of the stream, otherwise the peer would experience poor streaming quality. The incoming and outgoing bandwidth of peers is kept 512kbps that is equal to the streaming rate of the video. We deployed a single media source and the uplink bandwidth of media source is 20Mbs. The session width for push operation is 2 minutes. Whenever a peer receives the content from a parent peer in the tree, it keeps tracks of the sequence number of the packets. The peer in same session exchanges the sequence number as part of state information. For both data mining and popularity aware techniques, the playback records of each peer is evaluated. According to the record the playback procedure of a peer can be determined. For example a peer with playback record of (1, 2, 7, 9, 13, 14, 15, 19) has performed three seek operations in forward direction. The peers in both popularity aware and data mining based techniques request the popular content on the basis of this evaluation mechanism.

### 4.1. Result and Discussion

Fig. 5 shows the comparison of relative hit ratio for different prefetching techniques. *Relative Hit Ratio* is defined as the number of prefetching request satisfied locally. This means that the requested content, as a result of seek operation are already in local buffer of the peer. Both cooperative prefetching and data mining based technique have better hit ratio due to logical fetching of content. Cooperative prefetching focus on downloading segments near to the seek position, as well as rarest segments in the session. This ensures the availability of content in local buffer. On the other hand random prefetching has poor hit ratio due to blind prefetching of content without following any pattern.

Fig. 6 plots the global hit ratio for different prefetching techniques. We define "Global Hit Ratio" as the number of prefetching requests satisfied by requesting segments from neighboring peers.

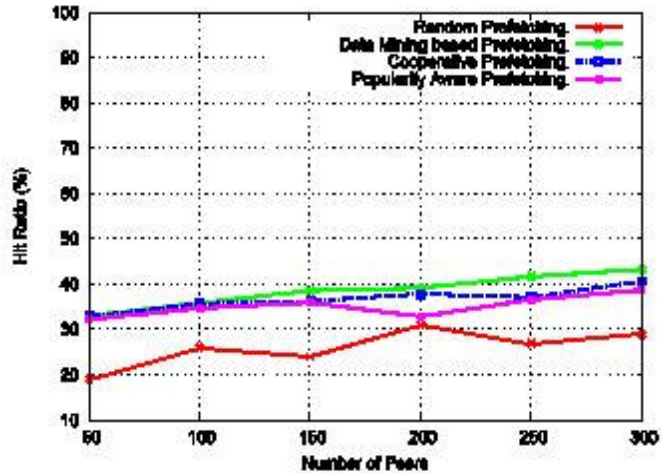

Figure 5: Comparison of Relative Hit Ratio

This means that requested segments are not available in local cache however they can be prefetched from other peers within the session.

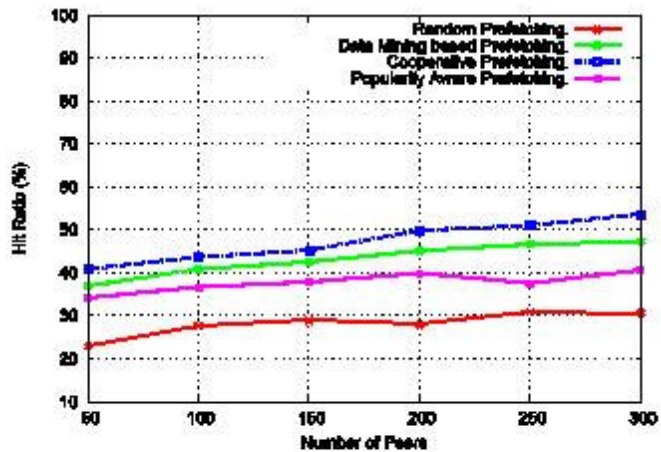

Figure 6: Comparison of Global Hit Ratio

Fig. 6 suggests that cooperative prefetching technique increases the overall hit ratio of the session. This is due to the reason that cooperative prefetching maximizes the availability of rare segments in the session. As a result maximum unique segments are available within the same session. Both data mining and popularity aware prefetching techniques focused on certain number of segments according to user behavior, while ignoring segments rarity.

Fig. 7 compares the response latencies for peers in different prefetching techniques. The result depicts that cooperative prefetching had lower response latency as compared to other existing prefetching mechanism. The peers in cooperative prefetching mechanism had lower interaction with far neighbors or server. The majority of the video content are available within the session. As a result peers obtained the content from

135

International Journal of Computer Networks & Communications (IJCNC), Vol.2, No.3, May 2010

their near neighbors and thus incur low response time. The response latency for early peers in cooperative prefetching mechanism is a bit higher. Once there are abundant resources in the session, the response latency tends to decrease.

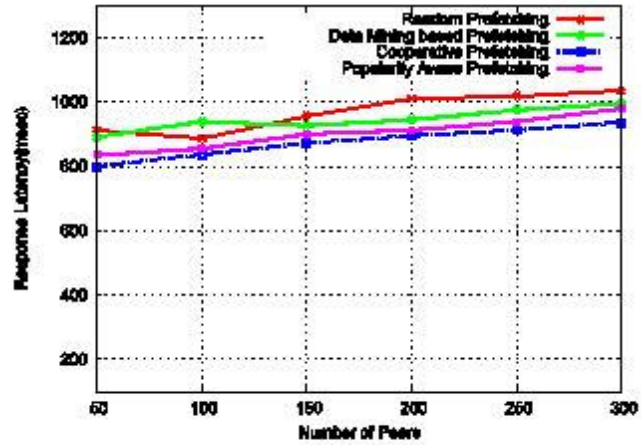

Figure 7: Comparison of response latencies

Fig. 8 and 9 shows the utilization ratio comparison of different prefetching techniques with cooperative prefetching.

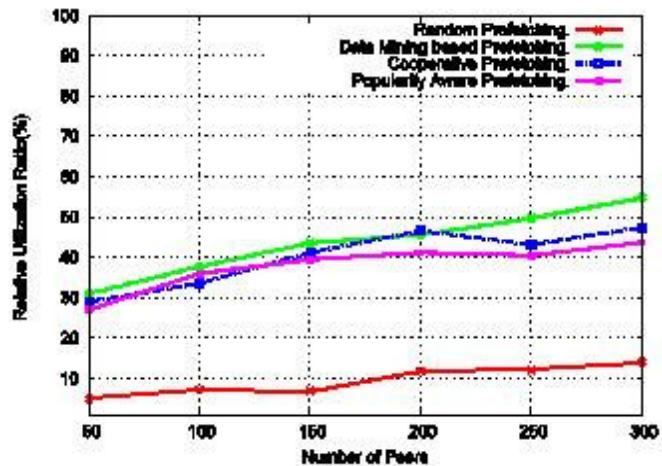

Figure 8: Comparison of relative utilization ratio

We defined "relative utilization ratio as the amount of prefetched data played to the total prefetched data cached at each peer". The "global utilization ratio" is the utilization ratio of the complete session. It is the ratio of utilized prefetched content to the total prefetched content available in the session. It is important to note that relative utilization ratio for cooperative prefetching is a bit lower than data mining based prefetching strategy as shown in Fig 8. This is understandable because our technique focus on prefetching rare segments into session and later on if any peer need a certain segment, it can prefetch from a peer in same session, with small delay. We observed that cooperative prefetching strategy utilizes the larger part of prefetched data for complete session as shown in Fig. 9, thus effectively utilizing the buffer.

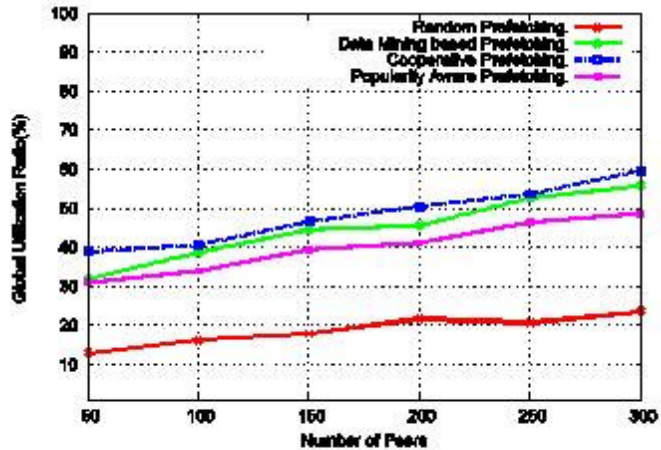

Figure 9: Comparison of global utilization ratio

The overhead comparison of different prefetching strategies is shown in Fig.10. We defined the overhead as control messages exchange for prefetching content. This includes the gossip messages and buffer map exchanges. It is observed that cooperative prefetching incurs less overhead as compared to popularity aware and data mining based strategy. Both popularity aware and data mining based techniques collects the user behavior data and then perform the necessary computations thus incurring greater overhead. In Fig. 10 it is observed that random prefetching had no overhead as it don't consider user behavior and fetch the content blindly, however it is not suitable due to poor hit ratio and utilization ratio.

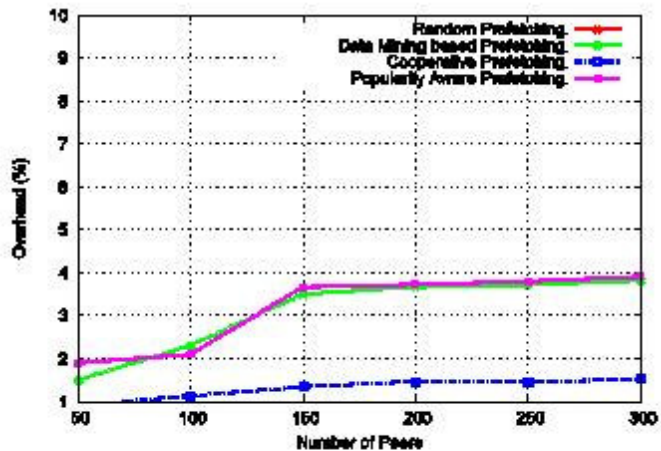

Figure 10: Overhead Comparison

## 6. CONCLUSION

In this paper we analyzed different existing prefetching techniques for peer assisted VoD system. In order to provide a VCR-oriented VoD service for P2P networks, we proposed a cooperative prefetching strategy. Our strategy focuses on improving the availability of rarest content in a session. Our proposed strategy improves the hit ratio and decrease the overhead significantly. Furthermore, we proposed a scheduling mechanism, which selects the best peer for provision of content. Finally, we believe that our results are promising and could provide research insight towards development of newer and efficient prefetching and caching strategies in P2P VoD systems. For the future perspective, we aim to perform real test-bed evaluation for the more personalized VoD and IPTV services delivery over P2P network.